\documentclass{llncs}
\begin{document}

\title{A Graph Spectral Approach for Computing Approximate Nash Equilibria}

\author{Haralampos Tsaknakis\inst{1} and Paul G. Spirakis\inst{1,2}}

\institute{Research Academic Computer Technology Institute (RACTI), Greece\\ Email: {\tt\{tsaknak,spirakis\}@cti.gr}
\and
Dept. of Computer Eng. and Informatics, Patras University, Patras Greece}

\newtheorem{theo}{Theorem}

\maketitle

\begin{abstract}
We present a new methodology for computing approximate Nash equilibria for two-person non-cooperative games based
upon certain extensions and specializations of an existing optimization approach previously used for the derivation of fixed approximations for this problem. In particular, the general two-person problem is reduced to an indefinite quadratic programming problem of special structure involving the $n \times n$ adjacency matrix of an induced simple graph specified by the input data of the game, where $n$ is the number of players' strategies. Using this methodology and exploiting certain properties of the positive part of the spectrum of the induced graph, we show that for any $\varepsilon > 0$ there is an algorithm to compute an $\varepsilon$-approximate Nash equilibrium in time $n^{\xi(m)/\varepsilon}$, where, $\xi (m) = \sum_{i=1}^m \lambda_i / n$ and $\lambda_1, \lambda_2, \ldots, \lambda_m$ are the positive eigenvalues of the adjacency matrix of the graph. For classes of games for which $\xi (m)$ is a constant, there is a PTAS. Based on the best upper bound derived for $\xi(m)$ so far, the worst case complexity of the method is bounded by the subexponential $n^{\sqrt{m}/\varepsilon}$.
\end{abstract}

\section{Introduction and notation}
It is well known that the problem of computing an exact Nash equilibrium is PPAD-complete even for $2$-player games.
Furthermore, the seemingly easier problem of computing an $\varepsilon$-approximate equilibrium for $\varepsilon$ inversely proportional to a polynomial in $n$ is also PPAD-complete. These results are established in \cite{chedete} and \cite{chede}. For the definition and insight into the complexity class PPAD and its connection with the more general Nash equilibrium problem, the interested reader is referred to \cite{papa} and \cite{DaGoPa}.

In view of such complexity results, attention has been focused in the past few years on the problem of finding
$\varepsilon$-approximate equilibria in polynomial time for some constant $\varepsilon$. However, despite considerable
efforts in this direction, it has not been possible so far to achieve in polynomial time guaranteed constant approximations better than $\varepsilon = 0.3393$ for general bimatrix games and $\varepsilon = 1/4$ for $\{0,1\}$ win-lose games. The latter approximation results are established in \cite{tsaspi} and \cite{tsaspi2} and have been achieved by using an optimization-based approach through the application of a descent algorithm that computes a stationary point of an
appropriately defined non-convex regret function. This optimization approach has also been experimentally tested and has been proven effective in practice, as demonstrated in \cite{tsaspikan}.

In this paper we are extending the latter approach by further exploring the special structure of the optimization
problem that represents a game. At first, the general problem of finding an approximate equilibrium of a bimatrix
game is reduced to a problem of finding an approximate symmetric equilibrium of a symmetric bimatrix game.  This
leads to an indefinite quadratic programming problem over the space of $n$-dimensional probability vectors involving
a single square nonnegative matrix $A$ and a regret function $f_A(x)$ to be minimized. The problem is further reduced
to a $\{0,1\}$ win-lose game, where, the matrix $A$ can be considered as the adjacency matrix of a simple directed graph,
called here the \textbf{induced graph} and which is uniquely specified by the game problem. Next, we consider a spectral
representation of the optimization problem in terms of the eigenvalues and eigenvectors of the adjacency matrix $A + A^\tau$ of the undirected version of the induced graph. The results presented here are based on the spectral properties of the graph in conjunction with the properties of stationary points of the regret function.

We adopt the following notation throughout the paper:

\begin{itemize}
  \item For any positive integer $k$, $[k]$ denotes the set of integers from $1$ to $k$. Also, $e$ denotes the
column vector of appropriate dimension having all its entries equal to $1$.
  \item ${\Delta}_k = \{ u : u \in \Re^k, u \geq 0, e^{\tau} u = 1\}$ is the $k$-dimensional standard simplex
(superscript $\tau$ denotes transpose). Alternatively, we refer to it as the set of $k$-dimensional probability vectors.
  \item $supp(u) = \{ i \in [k] : u_i \neq 0 \}$ denotes the subset of integers (indices) in $[k]$ constituting the
support of vector $u \in \Re^k$.
  \item $suppmax(u) = \{ i \in [k] : u_i \geq u_j ~\forall j \in [k] \}$ is the subset of indices in $[k]$ for which the corresponding entries of vector $u \in \Re^k$ are equal to the maximum entry value of $u$.
  \item $\max(u) = \{ u_i : u_i \geq u_j , \textrm{for all} j \}$ is the value of the maximum entry of vector $u$.
  \item $\max_{S}(u) = \{ u_i ,i \in S : u_i \geq u_j , \textrm{for all } j \in S \}$ is the value of the maximum
entry of a vector $u$ within an index subset $S \subset [k]$.
\end{itemize}

\section{Related results}

In regard to the complexity result derived from the methodology we present here, it is necessary to make a comparison with an existing subexponential algorithm for computing approximate Nash equilibria presented in \cite{LiMaMe}. This algorithm is based on an exhaustive search over all possible supports of size $O(\ln n / {\varepsilon}^2)$ that achieves an $\varepsilon$-approximate equilibrium in time $n^{O(\ln n / {\varepsilon}^2)}$. The proof of existence of the approximation is based on the probabilistic method. Considering as performance indicators the two parameters representing the approximation to a Nash equilibrium and the size of the game (i.e. the approximation parameter $\varepsilon$ and the number of strategies $n$ respectively), there is no uniform comparison between the complexity bound of the latter method and the one presented here: For a given $\varepsilon$, the result in \cite{LiMaMe} outperforms the result we present here for arbitrarily large $n$, larger than a threshold $n_0(\varepsilon)$. On the other hand, for a given $n$ our result outperforms the result in \cite{LiMaMe} for arbitrarily small $\varepsilon$, smaller than a threshold $\varepsilon_0(n) = n_0^{-1}(n)$. By taking into account also the constants involved in the exponents of the complexity bounds of the two methods (for the former method the constant multiplying $\ln n / {\varepsilon}^2$ is $12$ and for the method we present here the corresponding constant is less than $1$), the threshold curve is given approximately by the equation $\varepsilon \approx 12 \ln n / \sqrt{n}$. The comparison of the two results are shown below in tabular form ($m =$ number of positive eigenvalues $< n$):
\\\\
\begin{tabular}{|l|c|c|}
  \hline
   & Time & Approximation \\
  \hline
  New method & $n^{\sqrt{m}/\varepsilon}$ & $\varepsilon$ \\
  \hline
  Existing method & $n^{12 \ln n / {\varepsilon}^2}$ & $\varepsilon$ \\
  \hline
\end{tabular}
\\\\

It turns out that even for modest values of $\varepsilon$, the values of the threshold $n_0$ (above which the method of \cite{LiMaMe} outperforms our method) are simply too large, large enough to render any subexponential scheme totally unrealistic anyway.  Indicatively, for $\varepsilon = 1/3$ we have $n_0(1/3) \approx 2 \times 10^5$ and for $\varepsilon = 0.15$ we have $n_0(0.15)\approx 1.2 \times 10^6$ and this threshold increases very fast when $\varepsilon$ becomes even smaller. So, the method presented here is more efficient than the one presented in \cite{LiMaMe} for all practical purposes.

\section{Reduction to symmetric games}

Let $n$ be a positive integer and let $A$ be a real square $n \times n$ matrix. We assume, without loss of generality, that $A$ is normalized so that its maximum entry is $1$ and its minimum entry is $0$. For any $x \in \Delta_n$ we define the following function mapping $\Delta_n$ into $[0,1]$:

\begin{equation}
f_A(x) = \max(Ax) - x^\tau A x
\end{equation}

Minimizing $f_A(x)$ over $x \in \Delta_n$ is equivalent to finding symmetric Nash equilibria of the symmetric game $(A, A^\tau)$. It is also equivalent to finding Nash equilibria of the imitation game $(A, I)$, where $I$ is the $n \times n$ identity matrix.

Given a non-cooperative two-person game with payoff matrices $(R,C)$, the problem of finding an exact Nash equilibrium can be reduced to the problem of finding an exact symmetric equilibrium of a symmentric game, or, equivalently to finding a solution of an equation of the form $f_A(x) = 0$ for an appropriate choice of the matrix $A$.  The theorem below extends this result to approximate Nash equilibria as well.

\begin{theorem}
Given a procedure that for any (normalized) square $n \times n$ matrix $A$ and any $\epsilon > 0$ returns an $x \in \Delta_n$ such that $f_A(x) \leq \epsilon$, then, an $O(\epsilon)$-approximate equilibrium for any bimatrix game $(R,C)$ can be obtained in polynomial time.
\end{theorem}

\begin{proof}
Given a bimatrix game with payoff matrices $(R,C)$ (entries of both matrices are normalized in $[0,1]$), let us define the constants $c_1=\min_u \max_v (v^\tau R u)$ and $c_2=\min_v \max_u (v^\tau C u)$, where the $\min$ and $\max$ are taken over appropriately dimensioned probability vectors $v$ and $u$. Let $l_1$, $l_2$ be the number of rows and columns of the game respectively.

To avoid trivial cases we assume that no row of $R$ is dominated by its other rows (i.e. for every row of $R$ there is a $l_1$-dimensional probability vector whose inner product with this row is greater than its inner product with any other row of $R$) and, similarly, no column of $C$ is dominated by the its other columns. Indeed, if some row of $R$ (or some column of $C$) is dominated, then we can remove it from both matrices and reduce the dimensionality of the problem. In such cases, any Nash equilibrium of the reduced problem is also an equilibrium of the original game (notice that this procedure may eliminate some Nash equilibria of the original game. However, it will retain at least one). Also, we assume that no column of $R$ has all its entries equal to $0$ and no row of $C$ has all its entries equal to $0$. Indeed, if, for example a column $j$ of $R$ is zero, then we can easily construct a Nash equilibrium by having the column player play $j$ and the row player play a probability vector that makes $j$ dominant among the columns of $C$. A similar argument can be applied if a row of $C$ is zero. Note that checking for dominance of a row or column can be done in polynomial time by solving linear feasibility problems.

Using the above assumptions, it is easy to see that the constants $c_1$ and $c_2$ satisfy: $c_1 > 0$ and $c_2 > 0$.  Let us define the square $n \times n$ matrix $A$ (where, $n=l_1+l_2$) as follows :

\begin{displaymath}
A =
 \left[ \begin{array}{ccc}
0  &&  C^\tau \\
R  &&  0
\end{array} \right]
\end{displaymath}

Fix $\epsilon > 0$ such that $\epsilon < \frac{1}{2} min(c_1, c_2)$.  Let $x$ be an $n$-dimensional probability vector such that $f_A(x) \leq \epsilon$. Let $x_C$ and $x_R$ be the nonnegative vectors consisting of the first $l_2$ entries and the last $l_1$ entries of $x$ respectively, i.e. $x^\tau = (x_C^\tau, x_R^\tau)$. Under the specified constraints, it can be verified that $x_C \neq 0$ and $x_R \neq 0$.  Normalize $x_C$ and $x_R$ to make each one of them a probability vector (we use the same notation for simplicity). Let $M_1 = \max(R x_C)$ and $M_2 = \max(C^\tau x_R)$. Let $f_R(x_R, x_C) = M_1 - x_R^\tau R x_C$ and $f_C(x_R, x_C) = M_2 - x_R^\tau C x_C$ be the two regret functions for the row and column player respectively. Selecting a new $n$-dimensional vector $x_1^\tau = \frac{1}{M_1+M_2}(M_2 x_C^\tau, M_1 x_R^\tau)$, it can be verified that $f_A(x_1) \leq f_A(x) \leq \epsilon$. Expressing $f_A(x_1)$ in terms of $f_R$ and $f_C$ and using the inequalities $c_1 \leq M_1 \leq 1$ and $c_2 \leq M_2 \leq 1$, we obtain:

\begin{displaymath}
f_R (x_R, x_C) + f_C (x_R, x_C) \leq \frac{(M_1+M_2)^2}{M_1 M_2} \epsilon \leq 2 (\frac{1}{c_1} + \frac{1}{c_2}) \epsilon
\end{displaymath}
which proves the claim.

\qed
\end{proof}

Based on the above result we can focus our attention (without loss of generality) on the problem of approximating a minimum of a function of the form $f_A(x)$ for square matrices $A$ having the following properties:
\begin{itemize}
  \item All entries are nonnegative in the interval $[0,1]$
  \item Zero entries along the principal diagonal: $A_{ii} = 0$
  \item No column of $A$ is identically zero
  \item No row of $A$ is dominated by the other rows
\end{itemize}

\section{Stationary points and properties}
Following the analysis, results and terminology of \cite{tsaspi} and \cite{tsaspi2}, adapted here for the case of symmetric games (which makes things simpler, at least in terms of notation), we define stationary points of the function $f_A(x)$ and summarize their properties in a series of definitions and theorems below which we state without proof.

\begin{definition}
The \textbf{gradient} $D_A(x^\prime, x)$ of the function $f_A(x) = \max(Ax) - x^\tau A x$ at $x \in \Delta_n$ along a direction $x^\prime \in \Delta_n$ is defined as follows:
\begin{displaymath}
D_A(x^\prime, x) \equiv \lim_{\epsilon \to 0}\frac{1}{\epsilon} (f_A((1 - \epsilon) x + \epsilon x^\prime) - f_A(x))
\end{displaymath}
\end{definition}

\begin{theorem}
The limit in the above definition always exists for any $x \in \Delta_n$ and $x^\prime \in \Delta_n$. Furthermore, for given $x$ it defines a convex function $D_A(x^\prime, x)$ of $x^\prime$, involving the $\max(.)$ of an affine function of $x^\prime$, given by the equation:
\begin{displaymath}
D_A(x^\prime, x) = \max_{suppmax(A x)} (A x^\prime) - x^\tau A x^\prime - (x^\prime)^\tau A x + x^\tau A x - f_A(x)
\end{displaymath}
\end{theorem}

\begin{definition}
A probability vector $x^\star \in \Delta_n$ is called a \textbf{stationary point} of the function $f_A(x) = \max(Ax) - x^\tau A x$, if $D_A(x, x^\star) \geq 0, \forall x \in \Delta_n$.
\end{definition}

\begin{theorem}
A stationary point always exists and can be approximated as closely as desired in polynomial time through an iterative descent algorithm applied to the function $f_A(x)$.  The algorithm can start from an arbitrary $x_0 \in \Delta_n$ and involves solving a linear programming problem of the form $\min_{x \in \Delta_n} (D_A(x, x_k))$ at each step $k$.
Furthermore, every Nash equilibrium is a stationary point.
\end{theorem}

\begin{definition}
For any probability vector $x \in \Delta_n$, a probability vector $w \in \Delta_n$ is called an associated \textbf{dual} vector if it is a solution of the dual of the linear problem $\min_{x^\prime \in \Delta_n} (D_A(x^\prime, x))$, where all constraints are dualized except those specified by $x^\prime \in \Delta_n$.
\end{definition}

\begin{theorem}
For any probability vector $x \in \Delta_n$, the support of an associated dual vector $w \in \Delta_n$ satisfies $supp(w) \subseteq suppmax(A x)$.
\end{theorem}

\begin{theorem}
Let $x^\star \in \Delta_n$ be a stationary point and define $S(x^\star) \equiv suppmax(A x^\star)$. Let $w^\star \in \Delta_n$ be an associated dual probability vector (which satisfies $supp(w^\star) \subseteq S(x^\star)$ according to the previous theorem). Then, the following relationships (called here \textbf{stationarity conditions}) hold:
\begin{displaymath}
\max_{S(x^\star)} (A x) - (x^\star)^\tau A x - x^\tau A x^\star + (x^\star)^\tau A x^\star - f_A(x^\star) \geq 0, \forall x \in \Delta_n
\end{displaymath}
\begin{displaymath}
(w^\star)^\tau A x - (x^\star)^\tau A x - x^\tau A x^\star + (x^\star)^\tau A x^\star - f_A(x^\star) \geq 0, \forall x \in \Delta_n
\end{displaymath}
\end{theorem}

The first relationship in the above theorem yields:
\begin{equation}
f_A(x^\star) - f_A(x) + (\max (A x) - \max_{S(x^\star)} (A x)) \leq (x - x^\star)^\tau A (x - x^\star),  \forall x \in \Delta_n
\end{equation}

Also, setting $x = w^\star$ in the second relationship of the above theorem we obtain:
\begin{equation}
f_A(x^\star) \leq (w^\star - x^\star)^\tau A (w^\star - x^\star)
\end{equation}
or, equivalently, since $supp(w^\star) \subset S(x^\star)$:
\begin{equation}
2 f_A(x^\star) + f_A(w^\star) \leq \max(A w^\star) - (x^\star)^\tau A w^\star
\end{equation}

From the above equation we get:

\begin{theorem}
Given a stationary point $x^\star$ and associated dual $w^\star$, either $f_A(x^\star)$ or $f_A(w^\star)$ should be $\leq \frac{1}{3}$.
\end{theorem}

Another result which is a direct consequence of previous relationships involving stationary points is summarized in the theorem below.
\begin{theorem}
For any two stationary points $x_1^\star, x_2^\star$, the following relationship holds:
\begin{equation}
|f_A(x_1^\star) - f_A(x_2^\star)| \leq (x_1^\star - x_2^\star)^\tau A (x_1^\star - x_2^\star)
\end{equation}
\end{theorem}

As we will see later, equation $(5)$ above is important as it provides an upper bound on the quality of approximate Nash equilibria in terms of a symmetric quadratic form of the difference between stationary points. Notice that a similar relationship holds more generally for any pair of points $x_1, x_2 \in \Delta_n$ satisfying the condition that neither $x_1$ defines a descent direction for $f_A(.)$ at $x_2$ nor $x_2$ defines a descent direction for $f_A(.)$ at $x_1$, or equivalently, both $D_A(x_1, x_2)$ and $D_A(x_2, x_1)$ are $\geq 0$. If one of the two points, say $x_1$, is a Nash equilibrium and at the same time it does not define a descent direction for $f_A(.)$ at the other point $x_2$, then, $f_A(x_2) \leq (x_1 - x_2)^\tau A (x_1 - x_2)$. So, if any point $x_2$ is close to a Nash equilibrium in the squared metric defined by the symmetric quadratic form, then, it will either be an approximate equilibrium of the same quality, or there will be a descent direction from it. The above remarks are being used later in Section $7$ to establish the approximation result.

\begin{definition}
For a given convex set $K \subset \Delta_n$, a probability vector $x^\star \in K$ is called a \textbf{constrained stationary point} of the function $f_A(x) = \max(Ax) - x^\tau A x$, if $D_A(x, x^\star) \geq 0, \forall x \in K$.
\end{definition}
A constrained stationary point can be obtained in a similar way by starting from any $x_0 \in K$ and solving a linear programming problem of the form $\min_{x \in K} (D_A(x, x_k))$ at each step $k$ until convergence. In other words, we have additional constraints to deal with at each step of the descent algorithm. It is pointed out here that the rate of convergence of the descent algorithm to a (constrained) stationary point is not essentially affected by the additional constraints. The additional constraints affect the descent direction while the rate of convergence depends primarily on the stepsize given a descent direction. This can be verified by an overview of the convergence analysis presented in \cite{tsaspi} and \cite{tsaspi2}. Therefore, a constrained stationary point can also be obtained in polynomial time.

\section{Spectral representation}
Consider the spectral representation of the matrix $A + A^\tau$. Since this matrix is symmetric, all its eigenvalues are real and the eigenvectors are mutually orthogonal. Let $\lambda_1, \lambda_2, \ldots, \lambda_m$ be the $m$ positive eigenvalues of $A + A^\tau$ ($m < n$) and $-|\lambda_{m+1}|, -|\lambda_{m+2}|, \ldots, -|\lambda_n|$ the negative ones.  Let $z_i, i = 1, 2, \ldots, n$ be the corresponding normalized eigenvectors satisfying $\|z_i\| = 1$ for all $i \in [n]$ and $z_i^\tau z_j = 0$ for all $i, j$ such that $i \neq j$. By $\|.\|$ we denote the usual Euclidean $L_2$ norm. Assume that the eigenvalues are indexed in descending order, i.e. $\lambda_1 \geq \lambda_2 \geq \ldots \geq \lambda_n$.

Considering the representation of $A + A^\tau$ in terms of its eigenvalues and eigenvectors
\begin{equation}
A + A^\tau = \sum_{i=1}^m \lambda_i z_{i} z_{i}^\tau - \sum_{j=m+1}^n |\lambda_j| z_{j} z_{j}^\tau
\end{equation}
the function $f_A(x)$ can be written as follows:
\begin{equation}
f_A(x) = \max(Ax) + \frac{1}{2} \sum_{j=m+1}^n |\lambda_j| (z_{j}^\tau x)^2 - \frac{1}{2} \sum_{i=1}^m \lambda_i (z_{i}^\tau x)^2
\end{equation}

Let $A_{+} = \sum_{i=1}^m \lambda_i z_{i} z_{i}^\tau$ be the symmetric nonnegative definite $n \times n$ matrix of rank $m$ that defines the last term of the above equation involving the $m$ positive eigenvalues. The complexity of finding a solution of the problem $\min_{x \in \Delta_n}{f_A(x)}$ is due exclusively to this term since this term is responsible for the non-convexity of the problem. In particular, the complexity is driven by the geometry of the orthogonal projection of the feasible region of probability vectors $\Delta_n$ onto the $m$-dimensional linear subspace spanned by the eigenvectors $z_i, i = 1, 2, \ldots, m$ (that correspond to the positive eigenvalues) endowed with a distance metric $d(., .)$ defined by $d^2(a,b) = \sum_{i=1}^m \lambda_i (z_{i}^\tau (a - b))^2$, i.e. the metric induced by the matrix $A_{+}$. Expressing the distance between stationary points in terms of this metric, we notice that equation $(5)$ implies that for any two stationary points $x_1^\star, x_2^\star$ the following is true:
\begin{equation}
|f_A(x_1^\star) - f_A(x_2^\star)| \leq \frac{1}{2} d^2(x_1^\star,x_2^\star)
\end{equation}

We notice that the problem can be treated as an indefinite quadratic programming problem with linear constraints. A general result for such problems, obtained from \cite{vavasis}, states that there is an algorithm to find an $\epsilon$-approximate solution in $O \Big( K(n) \big(\frac{n(n+1)}{\sqrt{\epsilon}} \big)^m \Big)$ steps, where $K(n)$ denotes the time to solve a convex quadratic problem of size $n$.  An immediate consequence is that if the number $m$ of positive eigenvalues of $A + A^\tau$ is a small number, then there is an FPTAS for finding a Nash equilibrium. However, in such cases (where the number $m$ of positive eigenvalues is fixed) there is more structure in a game problem that can be exploited: Notice that $m$ is bounded by the rank of the matrix and as pointed out in \cite{ye} if the rank is fixed then there is a strongly polynomial algorithm to compute a Nash equilibrium.

The special structure of the game problem can generally be exploited to yield better results than the ones that can be obtained for an arbitrary indefinite quadratic programming problem. As we will see, it is not only the special properties of the feasible region (the set of $n$-dimensional probability vectors) that can be exploited, but also the properties of the eigenvalues and eigenvectors of nonnegative matrices and, more specifically, the spectral properties of adjacency matrices of simple graphs. The following theorem, known as the \textbf{Perron-Frobenius} theorem, states some basic facts about the principal eigenvalue and eigenvector of a nonnegative matrix:

\begin{theorem}
Any nonnegative square matrix has a nonnegative real eigenvalue $\lambda_1$ with maximum absolute value among all eigenvalues and a nonnegative real eigenvector $z_1$ corresponding to this eigenvalue. If the matrix has no block triangular decomposition, then $\lambda_1$ has multiplicity $1$ and the entries of the corresponding eigenvector $z_1$ are strictly positive.
\end{theorem}

In the sequel we will frequently make use of certain results of spectral graph theory, including the above mentioned theorem of Perron-Frobenius which is generally applicable to nonnegative matrices. For the basic theory and results of graph spectra the interested reader is referred to \cite{Bollo} and \cite{GoRo}.

In the next section we further explore the specific characteristics of a game as an indefinite quadratic programming problem in the context of win-lose $\{0,1\}$ games.  Considering win-lose games provides an explicit link of the game problem with the spectral properties of simple graphs.

\section{Win-lose games and induced graphs}
The analysis and results in the preceding sections are generally applicable to games with arbitrary payoffs. In this section we restrict attention to win-lose games where each payoff entry is either $0$ or $1$.

Considering win-lose games is no loss of generality in terms of the complexity of computing Nash equilibria. In \cite{AbbKaVa} it is proven that the computation of Nash equilibria for two-player games with rational payoffs is polynomially related to the computation of Nash equilibria for win-lose games, i.e. there is a polynomial time reduction of the Nash equilibrium problem from two-player games with rational payoffs to win-lose $\{0,1\}$ games. Therefore, finding a Nash equilibrium for $\{0,1\}$ games is also PPAD complete.

In order to avoid trivial cases (for example, cases where pure Nash equilibria exist), it is necessary to impose additional conditions on the matrix $A$ (in addition to those of Section 3): $A_{i j} + A_{j i} \leq 1$, i.e. that $A_{i j}$ and $A_{j i}$ cannot be both equal to $1$. Under this condition, the matrix $A$ can be considered as the adjacency matrix of a simple directed graph and, consequently, the matrix $A + A^\tau$ can be considered as the adjacency matrix of the undirected version of the same simple graph. We call this graph the \textbf{induced graph} of the game specified by the matrix $A$ and we denote it by $G = (V, E)$ with nodes the set of pure strategies $V = [n]$ and edges $E = \{(i, j): A_{i j} = 1\}$ $\subset V \times V$ the payoffs. Assume that the direction of each edge (in the directed version of the induced graph) is from a row to a column of matrix $A$, i.e. if $(i, j)$ is an edge, it is outgoing from $i$ and incoming to $j$.

Using the reduction to symmetric games obtained in Section 3 and the form of the matrix $A$ in the general case, we can consider only bipartite graphs without loss of generality. For bipartite graphs, the spectrum is symmetric, i.e. the number of positive eigenvalues is equal to the number of negative ones and each negative eigenvalue is equal in absolute value with a positive eigenvalue. However, the derivations and results presented here do not make explicit use of the bipartite nature of the underlying graph.

Now, in order to avoid other less trivial cases (for example, cases where the problem can be reduced to smaller dimension by a polynomial procedure) another condition on matrix $A$ should be imposed, if we take into consideration the Perron-Frobenius theorem, stated in the previous section. Specifically, the existence of a block triangular decomposition of the adjacency matrix implies that the induced graph $G$ is disconnected. In such a case we can decompose the game into subgames of smaller dimension, where, each subgame is represented by a principal submatrix of $A$ (a sumbatrix formed by a subset of rows and the corresponding subset of columns of $A$). Evidently, if we pick any subgame and find a Nash equilibrium of it, then, this equilibrium will also be a Nash equilibrium for the original game. Therefore, we assume that the game specified by $A$ cannot be decomposed, or, equivalently, the induced graph is connected (if this is not true, then we reduce the problem to lower dimension). Notice that we can check in polynomial time whether or not a graph is connected by solving a min-cut problem. Furthermore, as a consequence of the Perron-Frobenius theorem, if the induced graph is connected then the largest eigenvalue of $A$ has multiplicity $1$ and the corresponding eigenvector is strictly positive.

Based on the above remarks, we summarize below the assumptions that we make on matrix $A$, without loss of generality as far as win-lose games are concerned:

\begin{enumerate}
  \item Each entry of $A$ is either $0$ or $1$ and every column of $A$ contains at least one $1$.
  \item The entries of the principal diagonal are all $0$: $A_{ii} = 0, \forall i \in [n]$.
  \item The entries $A_{i j}$ and $A_{j i}$ cannot be both equal to $1$, i.e. $A_{i j} + A_{j i} \leq 1, \forall i, j \in [n], i \neq j$.
  \item The set of neighbors of any row of $A$ (the positions of $1$'s in the row) is no subset of the set of neighbors of any other row.
  \item The undirected graph with adjacency matrix $A + A^\tau$ is connected.
\end{enumerate}

As a corollary of the preceding discussion, we can express the following theorem:

\begin{theorem}
The problem of computing a Nash equilibrium for any bimatrix game can be polynomially reduced to a problem of computing a minimum of a function of the form $f_A(x) = \max(Ax) - x^\tau A x$, where $A$ is the adjacency matrix of a simple directed graph having properties $1-5$ above.
\end{theorem}
Recall that in case where any of the properties $1-5$ is not satisfied, the problem can be either reduced polynomially to a smaller one or there will be trivial Nash equilibria.

\section{Approximating an equilibrium}
In order to approximate an equilibrium, one can take advantage of the properties of stationary points (in particular equation $(5)$) as well as the spectral representation of the function $f_A(x)$ and the properties of the eigenvalues and eigenvectors of simple graphs. In principle, a grid of points can be appropriately constructed and using each point in the grid as a starting point, a stationary point can be computed using the descent algorithm mentioned in Section 4. If the grid is dense enough in a well defined sense, then, one of the stationary points produced should be close to an equilibrium. The important issue is how close one can get as a function of the price that has to be paid in terms of the density of the grid (hence the complexity of the algorithm).

Consider the linear $m$-dimensional subspace $\in \Re^n$ spanned by the eigenvectors $z_i, i = 1, 2, \ldots, m$ that correspond to the positive eigenvalues and the metric $d(., .)$: $d^2(a,b) = \sum_{i=1}^m \lambda_i (z_{i}^\tau (a - b))^2$ defined on this subspace. Consider the orthogonal projection of the feasible region $\Delta_n$ on this subspace and denote it by $P_m(\Delta_n)$. Denote by $P_m(x)$ the projection on this subspace of an $x \in \Delta_n$. It can be shown that $P_m(\Delta_n)$ is the convex hall of the projections of all the $n$ vertices of $\Delta_n$ on the subspace. In general, the number of vertices of the convex hall should be less than or equal to $n$. Notice that the projection of any probability vector in $\Delta_n$ on the subspace is an $m$-dimensional vector that can be expressed as a convex combination of at most $m+1$ vertices of the convex hall (by Caratheodory's theorem). Also, notice that the vertices of the convex hall can all be computed in polynomial time.

Let $\varepsilon$ be a positive approximation parameter and assume that $1/\varepsilon$ is integer smaller than $m$. Let us consider a set of regions in $P_m(\Delta_n)$ each consisting of convex combinations of no more that $1/\varepsilon$ vertices of $P_m(\Delta_n)$. Since each vertex is the projection of some vertex of $\Delta_n$, the set of such regions consist of the projections of all $n$-dimensional probability vectors with support no more than $1/\varepsilon$. The total number of such subsets of vertices of $P_m(\Delta_n)$ is $\leq n^{1/\varepsilon}$ and so is the total number of the corresponding regions. Let us denote the set of all such regions by $H(\varepsilon)$. This set of regions thus constructed constitute an approximation architecture of all points in $P_m(\Delta_n)$. The crucial question is how well does it approximate the entire $P_m(\Delta_n)$, i.e. what is the largest distance of a point in $P_m(\Delta_n)$ from the union of all sets in $H(\varepsilon)$ in the metric $d(., .)$. In regard to this question we can express the following theorem:
\begin{theorem}
For any $y \in \Delta_n$, the closest, in the metric $d(., .)$, point $x \in \Delta_n$ whose projection belongs to $\bigcup H(\varepsilon)$ satisfies the relationship: $d^2(P_m(x), P_m(y)) \leq \varepsilon \xi (m)$, where,
$\xi (m) = \sum_{i=1}^m \lambda_i / n$. \end{theorem}

\begin{proof}
Consider the matrix $A_{+} = \sum_{i=1}^m \lambda_i z_{i} z_{i}^\tau$. This is a nonnegative definite symmetric $n \times n$ matrix with rank $m$ which represents the positive part of the spectrum of $A + A^\tau$. For a given $y \in \Delta_n$ with large support (of the order of $m$ (but no more than $m+1$) to ensure that its projection $P_m(y)$ is as far as possible from any convex combination of a subset of $1/\varepsilon$ vertices of $P_m(\Delta_n)$), and $x \in \Delta_n$ such that $P_m(x) \in \bigcup H(\varepsilon)$ (i.e. the projection $P_m(x)$ of $x$ is a convex combination of no more than $1/\varepsilon$ vertices of $P_m(\Delta_n)$), we should have the relationship:
\begin{equation}
d^2(P_m(x), P_m(y)) = (x - y)^\tau A_{+} (x-y)
\end{equation}
Let the scalars $\bar{z_i}$ for $i = 1, 2, \ldots, n$ be defined as $\bar{z_i} = z_i^\tau y$. Define a new symmetric nonnegative definite matrix $A_{+}^\prime = \sum_{i=1}^m \lambda_i (z_{i} - \bar{z_i} e) (z_{i} - \bar{z_i} e)^\tau = (I - e y^\tau) A_{+} (I - y e^\tau)$ ($e$ is the all $1$'s vector). Then, given that $x$ and $y$ are probability vectors we can write the previous equation as:
\begin{equation}
d^2(P_m(x), P_m(y)) = x^\tau A_{+}^\prime x
\end{equation}
Let $\mu_1, \mu_2, \ldots $ be the eigenvalues of $A_{+}^\prime$. By construction, the sum of $\mu_i$'s (that is to say the trace of the matrix $A_{+}^\prime$) should be $\leq$ than $\sum_{i=1}^m \lambda_i$ (the trace of matrix $A_{+}$).
Notice that the set $supp(x)$ can be any subset of size $|supp(x)| \leq 1/\varepsilon$. So, the minimum of $d^2(P_m(x), P_m(y))$ with respect to $x$ is over all principal submatrices of $A_{+}^\prime$ of size $1/\varepsilon \times 1/\varepsilon$. Let $S(\varepsilon)$ be a subset of indices in $(1, n)$ defining such a submatrix and let $G(\varepsilon)$ be the submatrix itself. Then, $d^2(P_m(x), P_m(y)) = x^\tau G(\varepsilon) x$. It can be verified that, given an $S(\varepsilon)$, the minimum of the latter expression with respect to $x \in \Delta_n$ with $supp(x) \subset S(\varepsilon)$ is given by an expression of the form $1 / e^\tau G_{\varepsilon}^{-1} e$, where, $e$ here is the all $1$'s vector with support $S(\varepsilon)$ and $G_{\varepsilon}^{-1}$ is the inverse of a principal submatrix of $A_{+}^\prime$ of size $1/\varepsilon \times 1/\varepsilon$ if it exists, or it can be replaced by the pseudo inverse (generalized inverse) without loss of generality. It turns out that $d^2(P_m(x), P_m(y))$ can be bounded by an expression of the form $1 / \sum_{i \in S(\varepsilon)} \frac{1}{\mu_i^\prime}$, where, $\mu_i^\prime, i \in S(\varepsilon)$ are the eigenvalues of the submatrix $G(\varepsilon)$. Using the harmonic-arithmetic mean inequality, the latter expression is bounded from above by $\sum_{i \in S(\varepsilon)} \mu_i^\prime / |S(\varepsilon)|^2 = \varepsilon \sum_{i \in S(\varepsilon)} \mu_i^\prime / |S(\varepsilon)| = \varepsilon tr(G_{\varepsilon}) / |S(\varepsilon)|$. Since all submatrices $G_{\varepsilon}$ of size $1/\varepsilon \times 1/\varepsilon$ are considered, there is one whose average trace is minimum, which implies $tr(G_{\varepsilon}) / |S(\varepsilon)| \leq tr(A_{+}^\prime) / n$. So, the bound becomes:

\begin{equation}
d^2(P_m(x), P_m(y)) \leq \varepsilon tr(A_{+}^\prime) / m \leq \varepsilon \sum_{i=1}^m \lambda_i / n = \varepsilon \xi(m)
\end{equation}

Finally, the claim of the theorem follows from the last relationship.

\qed
\end{proof}

Based on the above theorem, we have the following result:
\begin{theorem}
For any $\varepsilon > 0$, there is an algorithm to find an $\varepsilon$-approximate Nash equilibrium in time
$n^{\xi(m)/\varepsilon}$, where, $\xi (m) = \sum_{i=1}^m \lambda_i / n$.
\end{theorem}
\begin{proof}

Since all points $y$ in $P_m(\Delta_n)$ can be covered by balls of the form $d^2(x, y) \leq \varepsilon \xi(m)$, for $x \in H(\varepsilon)$, we can consider all $n^{1/\varepsilon}$ points of $H(\varepsilon)$ as starting points for the descent algorithm and compute constrained stationary points within each such ball. One of them will be $\varepsilon \xi(m)$-close to a Nash equilibrium.

Also, since the parameter $\varepsilon$ can be chosen arbitrarily, one can choose $\varepsilon / \xi(m)$ in its place to get an $\varepsilon$-approximation in $n^{\xi(m)/\varepsilon}$ time.

\qed
\end{proof}

As an immediate consequence of the above theorems, we can express the following:

\begin{theorem}
For a class of games for which the positive eigenvalues satisfy the relationship $\sum_{i=1}^m \lambda_i / n =$ constant, then there is a PTAS for this class.
\end{theorem}

We conclude with the theorem below that establishes a general upper bound on $\xi (m)$ for all instances of games and corresponding induced graphs.

\begin{theorem}
The following relationship holds:
\begin{equation}
\xi (m) = \sum_{i=1}^m \lambda_i / n \leq \sqrt{m}
\end{equation}
\end{theorem}

\begin{proof}

Since the $\lambda_i$'s are the eigenvalues of the adjacency matrix $A + A^\tau$ of the induced undirected graph,
the sum of squares of the $\lambda_i$'s is equal to the trace of the matrix $(A + A^\tau)^2$ which is equal to the
number of walks of length $2$ in the graph for each node, i.e. twice the number of edges which is $\leq
n (n-1) / 2$. Using this fact and standard inequalities we obtain:
\begin{equation}
(\sum_{i=1}^m \lambda_i)^2 \leq m (\sum_{i=1}^m \lambda_i^2) \leq m (\sum_{i=1}^n \lambda_i^2) \leq m n^2
\end{equation}
which implies that:
\begin{equation}
\sum_{i=1}^m \lambda_i / n \leq \sqrt{m}
\end{equation}
\qed
\end{proof}

As a result of the last theorem, an $\varepsilon$-approximate equilibrium can be computed in time bounded by
$n^{\sqrt{m}/\varepsilon}$, where $m$ is the number of positive eigenvalues of the matrix $A + A^\tau$.

\section{Discussion and future work}
In this work it was demonstrated that the general problem of computing an approximate Nash equilibrium of a bimatrix game can be reduced to a problem of computing an approximate symmetric equilibrium of a symmetric $\{0,1\}$ win-lose game involving a single $n \times n$ matrix $A$ which is the adjacency matrix of a simple directed graph. The complexity of the problem is due to the positive part of the spectrum of the symmetric matrix $A+A^\tau$ (which is the adjacency matrix of the undirected version of the graph), i.e. to the eigenspace corresponding to the positive eigenvalues $\lambda_1, \lambda_2, \ldots, \lambda_m$. In particular, the geometry of the projection of the $n$-dimensional simplex $\Delta_n$ onto the positive eigenspace (denoted here by $P_m(\Delta_n)$) plays an important role in the analysis. As analyzed in Section $7$, for any $\varepsilon > 0$, $P_m(\Delta_n)$ can be covered by balls of squared radius $\varepsilon \sum_{i=1}^m \lambda_i / n$ centered at certain grid points in $P_m(\Delta_n)$ whose number does not exceed $n^{1/\varepsilon}$. Based on this and using the properties of stationary points as summarized in Section $4$, it is shown that by running the descent algorithm at most $n^{1/\varepsilon}$ times independently, each starting from a distinct point of the grid, one of the stationary points thus obtained is $\varepsilon \sum_{i=1}^m \lambda_i / n$-close to an equilibrium. Finally, by appropriate choice of the parameters, it turns out that it is possible to achieve an $\varepsilon$-approximate equilibrium in time less than $n^{\xi(m)/\varepsilon}$, where $\xi (m) = \sum_{i=1}^m \lambda_i / n$.

There are some issues that can be further investigated to improve the results. At first, the grid that was used to subdivide $P_m(\Delta_n)$ basically consists of the projections on the positive eigenspace of all $n$-dimensional probability vectors whose entries are integer multiples of a given $\varepsilon$. The question is: Is it possible to further exploit the structure of $P_m(\Delta_n)$ and to use a different grid containing less points that could yield better results?

Secondly, and more importantly, the complexity as a function of the positive eigenvalues depends on the quantity $\xi (m) = \sum_{i=1}^m \lambda_i / n$ which is always $\leq \sqrt{m}$ (which implies our subexponential complexity result in this paper). However, the $\sqrt{m}$ bound for $\xi (m)$ is not tight. In fact, the equality holds only if all positive eigenvalues are equal (i.e. the multiplicity of $\lambda_1$ is $m$) which in turn implies that the induced graph is not connected (this is a consequence of the Perron-Frobenius theorem and a standard result in spectral graph theory). But if the graph is not connected, the game problem can be reduced in dimension as analysed in Section $6$. The question is: Is there a better upper bound in any significant way for $\xi (m)$ that explicitly takes into account the connectivity of the induced graph?

As pointed out in Theorem $12$, if for some class of games the expression $\xi (m)$ is a constant, then our approach leads to a PTAS for this class. In general, it appears that the study of the positive spectrum of the adjacency matrix $A+A^\tau$ can elucidate several aspects of an equilibrium problem (notice that eigenvalues and eigenvectors can be computed in polynomial time). The expression $\xi (m)$ can be used to characterize classes of game problems according to the difficulty in approximating equilibria.

The distribution of the principal graph eigenvalues and corresponding eigenvectors are already being heavily used in a variety of applications such as in expander graphs, spectral partitioning algorithms, page ranking algorithms, etc. It is conjectured that further study of the principal eigenvalues and eigenvectors of graphs with an eye on game problems can yield improved results and algorithmic approaches for the computation of approximate and exact game equilibria as well. \\

\textbf{Acknowledgements:} We wish to thank C. H. Papadimitriou, X. Deng and Y. Ye for many useful comments on an earlier version of this work.

\end{document}